\documentclass[lettersize,journal]{IEEEtran}

\usepackage{amsmath,amsfonts,amssymb}
\usepackage{algorithmic}
\usepackage{array}
\usepackage[caption=false,font=normalsize,labelfont=sf,textfont=sf]{subfig}
\usepackage{balance}
\usepackage{textcomp}
\usepackage{stfloats}
\usepackage{xfrac}
\usepackage{url}
\usepackage{cite}
\usepackage{mathrsfs}
\usepackage{verbatim}
\usepackage{booktabs}
\usepackage{adjustbox}
\usepackage{graphicx}
\usepackage{xcolor}
\usepackage{soul}
\usepackage{framed}
\usepackage{multirow}
\usepackage{lipsum}

\usepackage[hidelinks]{hyperref}

\colorlet{shadecolor}{yellow}

\begin{document}

\title{Brain-Inspired User Positioning for mmWave Beam Management}

\author{%
Aris Karampelas Timotijevic,
Evangelos Koutsonas,
Vasileios Kouvakis,
Stylianos E. Trevlakis,
Alexandros-Apostolos A. Boulogeorgos, and Theodoros A. Tsiftsis
\thanks{This work was supported by the research project NEURONAS.
The research project NEURONAS is implemented in the framework
of H.F.R.I. call ``3rd Call for H.F.R.I.'s Research Projects
to Support Faculty Members \& Researchers''
(H.F.R.I. Project Number: 25726).}%
\thanks{A. Karampelas Timotijevic and T. A. Tsiftsis are with the
Department of Informatics and Telecommunications,
University of Thessaly, Lamia 35100, Greece
(e-mail: \texttt{\{akarampel, tsiftsis\}@uth.gr})
.}
\thanks{%
E. Koutsonas is with the Department of Electrical and
Computer Engineering, University of Western Macedonia,
50100 Kozani, Greece (email: \texttt{dece00106@uowm.gr}).%
}%
\thanks{%
V. Kouvakis and S. E. Trevlakis are with the Department
of Research and Development, InnoCube P.C.,
55534 Thessaloniki, Greece
(e-mail: \texttt{\{kouvakis, trevlakis\} @innocube.org}).}%
\thanks{%
A.-A. A. Boulogeorgos is with the Department of Electrical
and Computer Engineering, Democritus University of Thrace,
Xanthi, Greece
(e-mail: \texttt{abouloge@ee.duth.gr}).
}
}

\maketitle
\begin{abstract}
Wireless positioning is a key enabling functionality of high-frequency, highly directional wireless systems. Supporting beam tracking depends on accurately translating channel state information (CSI) estimations into position predictions. However, as urban propagation environments are inherently highly complex, determining the process that connects the CSI with the mobile user position is a difficult task. Motivated by this observation, a great amount of effort was put on designing user positioning schemes that employ conventional machine learning approaches, such as $k$-nearest neighbors ($k$-NN), and convolutional neural networks (CNNs). However, conventional models achieve low energy efficiency (EE). To counterbalance this disadvantage, brain-inspired neural processing units (NPUs) has recently been introduced. For their optimum operation, NPUs require the execution of a new type of neural models, namely spiking neural networks (SNNs). Inspired by this, in this paper, we present a wireless network architecture that enables CSI acquisition, translation into mobile user position information, and beamforming adaptation through novel convolutional SNN (CSNN) models. To train and assess the models' performance, we created datasets based on realistic ray-tracing-based simulations. We applied the CSNN architecture and compared its performance against fingerprinting $k$-NNs, and CNNs in terms of both computing- and communication-tailored performance metrics. The results highlight a trade-off between EE and user positioning accuracy. 
\end{abstract}


\section{INTRODUCTION}
\IEEEPARstart{P}{ositioning} is a fundamental functionality in modern wireless systems \cite{Italiano2025}. Accurate knowledge of user locations enables beamforming optimization; thereby, improving the system's reliability \cite{Kallehauge2024} and energy efficiency \cite{Koivisto2017}. In addition, positioning plays a key role in several killer applications, including autonomous transportation \cite{Mahmoud2020}, environment-aware networking \cite{Si2024}, search-and-rescue (SAR) operations \cite{Huang2022}, smart factories \cite{Xiao2025}, and extended reality \cite{Ma2023}.

Conventional wireless positioning methods commonly rely on temporal, angular, or power-related characteristics of the received signal \cite{Bensky2016}. Time-of-arrival (ToA) techniques estimate propagation delays between transmitted and received signals to infer distances, while angle-of-arrival (AoA) techniques determine the direction from which signals impinge on an antenna array. Another widely used approach exploits the received signal strength indicator (RSSI), where signal attenuation serves as a basis for coarse distance estimation \cite{Shibata2026}. In indoor environments, RSSI-based fingerprinting methods utilize pre-measured radio maps to estimate user positions from observed channel conditions \cite{Kawecki2022, Li2023Data}.

High-frequency systems that operate in the millimeter-wave (mmWave) or the terahertz (THz) bands, provide new opportunities for high-precision positioning \cite{Kanhere2018, Adnan2025}. Compared to conventional sub-6~GHz systems, high-frequency systems employ considerably larger bandwidths and highly directional beams, enabling improved spatial and temporal resolution. These characteristics allow positioning accuracy enhancements. On the other hand, high-frequency communications suffer from severe propagation losses, and sensitivity to blockage. Under such rapidly varying channel conditions, the additional signaling overhead associated with conventional positioning procedures may become impractical.

To address the highly nonlinear relationship between wireless channel characteristics and user position, machine learning (ML) and deep learning (DL) methods have recently emerged as promising solutions \cite{dichasus2021, Ismail2024, Utama2024, Respati2026}. These methods utilize input features such as RSSI, channel state information (CSI), AoA, and ToA to learn complex mappings between observed channel conditions and spatial coordinates. In particular, CSI-based learning approaches have demonstrated strong positioning performance due to the rich propagation information embedded within CSI measurements.

In more detail, the authors of \cite{dichasus2021} established an experimental indoor setup by using conventional transmitters and receivers. Such approaches are more feasible for indoor positioning systems due to the lower complexity of the propagation environment. 

Raw CSI data \cite{Uykan2026, Respati2026} or even RSSI measurements \cite{Shibata2026} can be fed directly to DL models, enabling the models to approximate the nonlinear relationship between CSI and user coordinates. An alternative approach is the use of ray-tracing models. CSI generation through simulations based on realistic ray-tracing channel models has gained significant traction in recent years \cite{sionna2022}.
 
Most datasets include several channel and propagation characteristics, thus allowing for evaluation of positioning systems under conditions in real-world setups.

Early deployment of ML and DL models primarily relied on cloud-based inference, where data were uploaded to centralized servers for processing \cite{Li2019}. This approach introduced significant communication overhead and inference latency, making it unsuitable for latency-sensitive wireless applications. Consequently, inference has progressively shifted toward edge devices, giving rise to the concept of edge artificial intelligence (AI) \cite{Shuvo2023, Wang2025, Ngo2025, Quadar2026}. Edge AI enables low-latency processing closer to the data source but simultaneously introduces strict constraints on computational complexity, energy consumption, and hardware efficiency. These challenges have motivated the development of AI-gnostic lightweight and energy-efficient brain-inspired computer architectures \cite{Lin2024}.
 
Brain-inspired architectures, such as neuromorphic units, provide an event-driven and energy-efficient computing paradigm. In contrast to conventional digital hardware, which relies on continuous signal processing and repeated memory-access operations, neuromorphic systems exchange information through discrete spike events among computational units. As a result, computation and communication are activated only when meaningful events occur, which can substantially reduce energy consumption. 
 
The machine learning implementation is typically realized through SNNs, which model information processing using spike-based neurons, such as integrate-and-fire (IF) and leaky integrate-and-fire (LIF) models. 

By organizing such neurons into multilayer architectures, SNNs can perform inference on complex tasks, including classification \cite{Eshraghian2021, Hu2024} and regression \cite{Henkes2024, Happek2024, Huang2025}. 

SNNs have been applied to integrated sensing and communications, where a single neuromorphic receiver can simultaneously perform wireless data decoding and radar target detection using impulse-radio signals, improving energy efficiency and supporting low-latency operation for future IoT and 6G networks \cite{Chen2023}. 
In non-terrestrial networks (NTNs), SNNs are applicable to onboard inference for airborne and spaceborne payloads. In \cite{Eappen2025}, an interference detection and identification brain-inspired model is proposed. Through experimental evaluation, the energy efficiency advantage of SNNs over a conventional CNN is demonstrated. Positioning using SNNs has been implemented using conventional mobile devices. By leveraging orientation and velocity measurements from sensors found on mobile devices, an SNN model can infer the user's position using regression \cite{Huang2025}. To the best of our knowledge, CSNN models for CSI-based positioning have not been investigated. Although existing works leverage CSNN models for image-based object recognition, no prior work has applied the CSNN model in telecommunications. As such, the application of a CSNN to CSI data, derived from ray-traced channel simulations, constitutes the main novelty of this work. 

Motivated by this, we propose a brain-inspired positioning sub-system for a mmWave system using a CSNN. Through our work, we study the highly non-linear relationship between CSI and user coordinates. In more detail, the technical contribution of this paper can be summarized as:
\begin{itemize}
\item We design and apply CSNN models that estimate the wireless user equipment (UE) position through CSI data  acquired at the uplink (UL).
\item We provide a comprehensive comparison of distinct neural models under a unified evaluation framework. Specifically, a unified benchmark evaluated under a single framework spanning both ML and communication metrics on identical data and  hardware.
\item We explicitly quantify the accuracy-energy trade off for neuromorphic evaluated positioning. The latter shows that CSNN occupies an intermediate accuracy position, while achieving the lowest estimated per-inference neuromorphic energy.
\end{itemize}

The rest of the paper is organized as follows. The system model is presented in section \ref{sec:sysmodel}. In section \ref{sec:nn}, we present the proposed positioning model, and benchmarks, with the performance metrics presented in the subsequent section \ref{sec:perf_metrics}. Section \ref{sec:results} outlines the simulation parameters and reports our findings. Finally, section \ref{sec:conclusion} presents the conclusions of our work.

\textbf{\textit{Notation:}}~Boldface (lower case) $\mathbf{x}$ represents vectors, (upper case) $\mathbf{X}$ denotes matrices, and scalars are indicated by letters, both upper and lower case. Let $\mathbf{X}^T, \mathbf{X}^H, \mathrm{diag}(\mathbf{X})$, and $\mathrm{tr}(\mathbf{X})$ represent the transpose, hermitian, diagonalization, and trace of $\mathbf{X}$, respectively. $X_{i,j}$ denotes the $(i,j)$-th entry of matrix $\textbf{X}$, while $x_i$ denotes the $i$-th entry of vector $\mathbf{x}$. Calligraphic (uppercase) $\mathcal{X}$ represent sets. For a set $\mathcal{X}$, $|\mathcal{X}|$ denotes the cardinality. 

\section{SYSTEM MODEL \& COMMUNICATION PROTOCOL }\label{sec:sysmodel}
    This section provides the construction of the system model, its technical characteristics, and the employed beamforming techniques. In addition, it presents a comprehensive communication protocol alongside the channel model.
\subsection{System Model}
\begin{figure}
        \centering
        \includegraphics[width=\columnwidth]{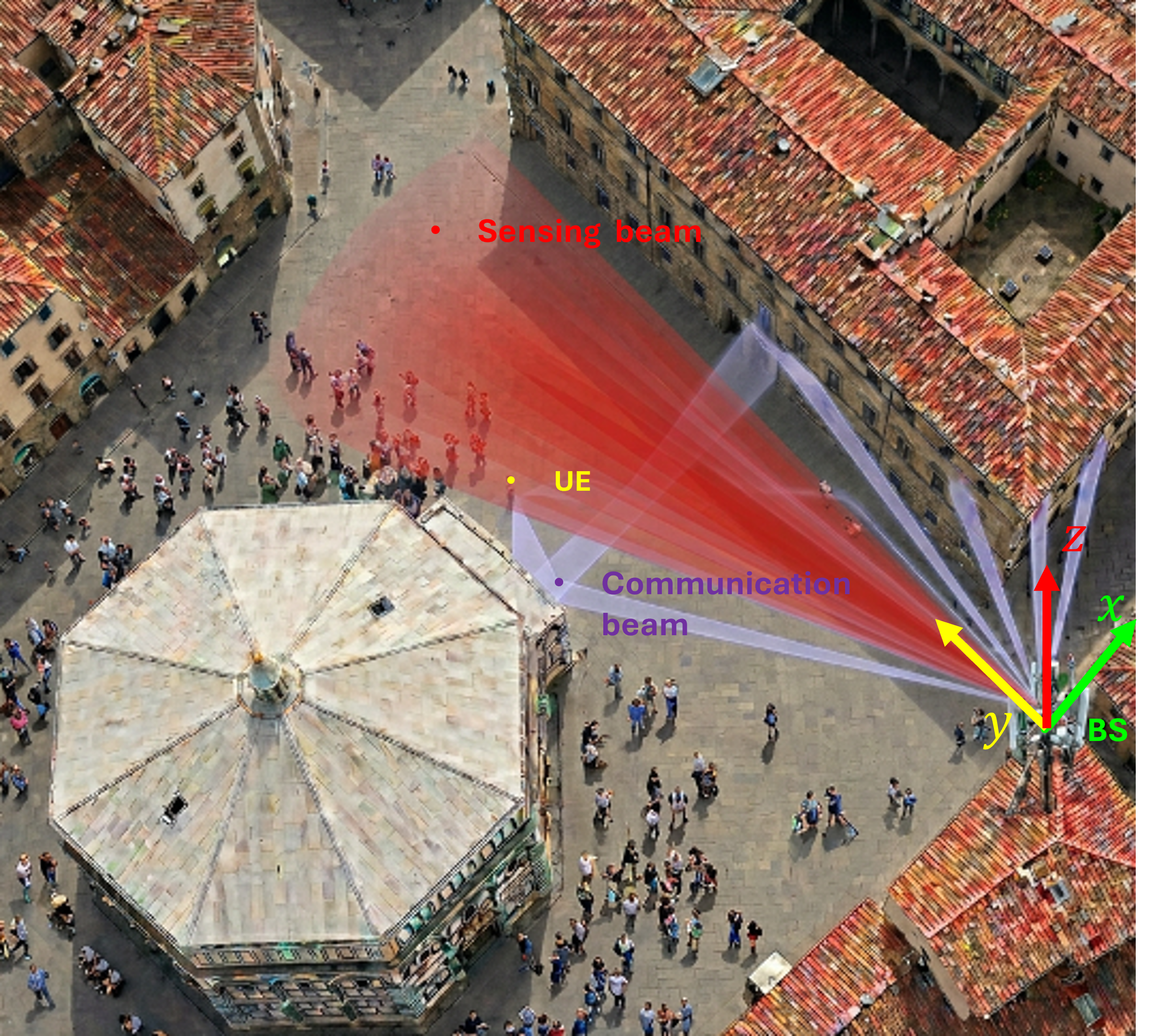}
        \caption{System model.}
        \label{fig:poi}
    \end{figure}

    As illustrated in Fig.~\ref{fig:poi}, we consider an urban simultaneous transmit and receive scenario, in which $K$ UEs communicate with a BS using orthogonal frequency division modulation (OFDM). The channel is expected to experience multipath fading, which can be accurately modeled through ray-tracing techniques. The BS is equipped with an uniform rectangular planar array (URPA) that consist of $N_{BS}=N_A\times N_B$  antenna elements, while each UE is equipped with a single antenna. To enable massive connectivity, while maintaining low-complexity, the BS employs a hybrid beamforming (HBF) with $N_{\rm{RF}}$ radio frequency (RF) chains, as described in \cite{partially_HBF}.

    At the BS, one RF chain is dedicated for sensing, whereas the remaining RF chains ($K_{\rm DL}$) are used for communication. 
    The sensing RF chain is connected to  $N_{RX}=N_{RX}^x\times N_{RX}^y$ antenna elements, where $N_{RX}^x$ and $N_{RX}^y$ stand for the horizontal and vertical elements, respectively. 
    
    In the UL, the BS operates as a receiver, so the combining vector  can be defined as 
    \begin{equation}
        \boldsymbol{\phi}=\left[\phi_1, \phi_2, \dots, \phi_{N_{RX}} \right]^T
    \end{equation}
    where $\phi_i=e^{j\phi_{S_i}}$ represents the phase shift associated with the $i$-th receiving antenna element, and $\phi_{S_i}$ denotes the corresponding beam steering phase angle.    
    Similarly, in the downlink, the BS operates in transmitter mode; thus, the combining diagonal matrix can be defined as
    \begin{align}\label{eq:FDL}
        \mathbf{F}_{\mathrm{DL}} &= \boldsymbol{\Psi}_{\mathrm{DL}}\,\mathbf{B}_{\mathrm{DL}} 
    \end{align}    
    where $\mathbf{\Psi}=\rm{diag}\left[ \Psi_1, \Psi_2, \ldots, \Psi_{N_{TX}}\right]$
     represents the diagonal phase shifter matrix associated with the transmitting antenna elements. $\Psi_i = e^{j\theta_{S_i}}$ denotes the phase shift associated with the $i$-th transmitting antenna element, where $\theta_{S_i}$ denotes the corresponding beam steering phase angle. Furthermore, $\mathbf{B}~=~\rm{diag}\left[ \mathbf{b}_1,\mathbf{b}_2, \ldots, \mathbf{b}_{N_{TX}}\right]$
    denotes the on/off state of the corresponding phase shifter with $\mathbf{b}_i$ representing an all-one column vector.
    The BS location is fixed and known a priori known to the $K$ UEs, whereas the UEs move randomly; thus, the UE locations are unknown to the BS. 
    The UEs are assumed to be located and moving within a predefined two-dimensional Cartesian coordinate region, defined as
    $\mathcal{A}~\triangleq~\left\{\left(x,\,y\right)\in\mathbb{R}^2\mid 0\leq x\leq X,\,0\leq y\leq Y\right\}$.
    
    \subsection{Communication Protocolol}
    The communication cycle is divided into three phases. During the first phase, once specific radio resources are allocated by the BS, an initial access signal is broadcast over the entire considered area. 
    In the second phase, the UEs 
    transmit an initial access request message to the BS. Then, the BS selects the UE to serve based on predefined precedence and performs UE position prediction. In the third and final phase, the BS establishes communication through a highly directional downlink beam. To maintain alignment, the BS continues identifies the position of the UE through the wide directional UL connection. 
    
    During the UL transmission phase, each UE periodically communicates with the BS. The communication duration is set to $\Delta t$. The link employs OFDM comprising $N$ subcarriers at a central frequency $f_{c}$ and subcarrier spacing (SCS) $\Delta f$.
    Thus, the signal received at the BS from the $k_{th}$ UE can be written as
    \begin{equation}
\boldsymbol{y_k}=\mathbf{F^{^H}_{UL}}\, \mathbf{h_k}\,+\,\boldsymbol{n_o}
         \label{ULsignal}
    \end{equation}
    where $\boldsymbol{n_o}$ represents a zero-mean additive white Gaussian noise (AWGN) with variance $\sigma^2_{_{RT}},$ and $\mathbf{h_k}$ represents the channel frequency response (CFR) of the ray-tracing channel given by \cite{Fugen2006}
    \begin{equation}
      \begin{aligned}
        \mathbf{h_k}= \sqrt{ {\left(\dfrac{\lambda_c}{4\,\pi}\right) }^{2}\,G_t\,G_r }\,\sum_{m=1}^{M}\,\mathbf{a}_m\,{e}^{-j2\pi f\tau_{m_{\mathrm{pd}}}(t)}   
      \end{aligned}
      \label{CFR}
    \end{equation}
    where $\lambda_c$ denotes the wavelength corresponding to the central frequency of the system, $c_0$ represents the speed of light, and $G_t$ and $G_r$ denote the transmitter and receiver antenna gains, respectively. Furthermore, $M$ represents the number of propagation paths, and $\tau_{m_{\mathrm{pd}}}$ is the propagation delay of the $m$-th path.
    Additionally,
    \begin{equation}
      \begin{aligned}
        \mathbf{a}_m= {\mathbf{C^H_{RX}}\left(\theta_m,\,\phi_m\right)}\mathbf{T_m}\,\mathbf{C_{TX}}\left(\theta_m,\,\phi_m\right)     \end{aligned}
    \end{equation}
     where, $\theta_m$ and $\phi_m$ are the angles of departure (AoD) and arrivals (AoA) of the $m$-th path, respectively, $\mathbf{T_m}$ denotes the polarimetric transmission matrix of the $m$-th path, while $\mathbf{C_{RX}}(\cdot)$ and $\mathbf{C_{TX}}(\cdot)$ return the receive and transmit antenna patterns, respectively. 

\section{PROPOSED POSITIONING MODELS}\label{sec:nn}

In this section, we present two neural network architectures for the wireless user positioning problem in the considered mmWave system. We also report a ML model as a baseline. We further analyze the computational complexity of each model and define the performance metrics used to assess their effectiveness.

\subsection{Data Processing for the Input Layer}\label{subsec:input_preproc}

As illustrated in Fig.~\ref{fig:poi}, we define an antenna-centric coordinate system. The antenna is placed at $\mathbf{p}_\mathrm{BS} = [0, 0, z]^T$, where $z$ is the antenna elevation. All the position estimates, $\hat{\mathbf{p}}$, and the ground-truth positions, $\mathbf{p}$, belong to this coordinate system.

We also apply a transformation to the CSI tensor. Each CSI tensor element is complex-valued. ANNs are not suited to processing complex-value inputs, so real-valued representation is required \cite{Paryanti2020}. To achieve this, we separate the real and imaginary parts of each CSI entry. Consequently, we extend the tensor with an additional dimension of size 2. The new dimension contains the real and imaginary parts for each CSI entry. The elements now belong to the real domain, and the shape of each tensor fed to the input layer is $[N_{RX},~N,~2]$.

To avoid feeding an extremely large number of features and increasing the input layer size excessively, we follow the approach provided in \cite{dichasus2021}. Specifically, we average adjacent subcarriers. We split the CSI tensor along the second dimension into chunks of size $N / p$, where $p$ is the chunk size. For each chunk, we compute the mean across that chunk, and then we stack all chunks into a tensor. This reduces the size of the input layer by a factor $p$.

\begin{figure*}[t]
\centering
\includegraphics[width=\linewidth]{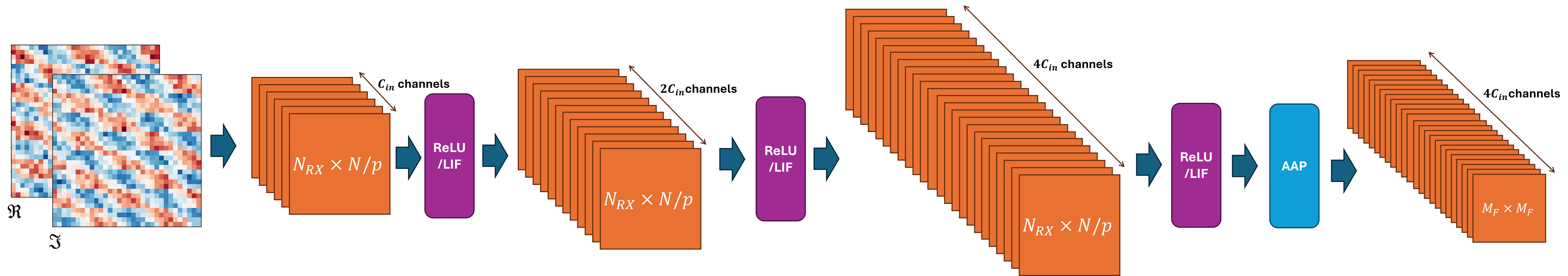}
\caption{Proposed feature extractor. The CSI measurements are fed to the feature extractor as two-channel images, with the channels being the $\Re\{\cdot\}$ and $\Im\{\cdot\}$ of the measurement. The conventional architecture uses ReLU layers between the convolutions, while the brain-inspired uses LIF layers. The final ReLU/LIF layer passes the features to an Adaptive Average Pooling (AAP) layer to reduce the dimensionality of the features.}
\label{fig:architectures}
\end{figure*}

\subsection{Benchmark Models}

In this section, we present the benchmark models.

\subsubsection{$k$-NN}

A fingerprint is a CSI measurement paired with the position it was taken. With the aggregation of enough measurements a radio map can be constructed. To position a new measurement, the $k$-nearest neighbours ($k$-NN) regression selects the $k$-stored fingerprints whose fingerprint most resembles the new measurement, and predicts a weighted average of their positions. Consequently, the radio map serves as an extensive look-up table (LUT), $\{ (\mathbf{h}_i,~\mathbf{p}_i) \}$. The new measurement is represented by a query CSI measurement, $\mathbf{h}_q$. Considering the Euclidean norm as the distance metric the position estimation is calculated as
\begin{equation}
    \hat{\mathbf{p}} = \frac{\sum_{i\in N_k}{w_i\mathbf{p}_i}}{\sum_{i\in N_k}{w_i}},
\end{equation}
where $N_k$ is the number of nearest fingerprints, $w_i=1/d_i$ is the fingerprint weight, and $d_i=|| \mathbf{h_q} - \mathbf{h_i} ||_2$ the Euclidean norm between the CSI query and the $i$-th nearest fingerprint.

\subsubsection{CNN}
CNNs are mainly used when the input data are multi-dimensional. An indicative example of multi-dimensional data is an image. CNNs rely on convolution to extract useful features from the multi-dimensional data. As it is widely known, convolution is basically a filter operation, where certain characteristics of an input tensor are extracted. 
A typical CNN architecture consists of two parts. The first part is called convolutional segment, and is tasked with approximating the convolution kernels required for a desired output. Between the convolution layers, there are several other layers, such as batch normalization layers (BatchNorm) and pooling layers, that contribute to the network training. Batch normalization layers enable the model's convergence, while pooling layers allow for spatial dimensions reduction. The output of the convolutional segment is called a feature map. The feature map is flattened and fed into a linear segment, which is basically an fully-connected network, and used for either regression or classification tasks.

In Fig.~\ref{fig:architectures} the architecture of the convolutional layers for feature extraction is presented. The input tensor is first rearranged into a channel-first representation before being processed by three successive convolutional layers with increasing channel depth, $C_\mathrm{in}$. Each convolutional block employs a $M_{\mathrm{K}} \times M_{\mathrm{K}}$ kernel followed by a ReLU activation function to extract hierarchical spatial features from the input measurements. An adaptive average pooling layer reduces the spatial dimensionality to a fixed $M_{\mathrm{F}}\times M_\mathrm{F}$ representation, enabling stable feature aggregation independent of the original input dimensions. The extracted feature maps are subsequently flattened and passed through a fully-connected regression head that estimates the final $(x,y)$ position coordinates. The regressor head employs a single hidden layer with $N_n$ neurons.

\subsection{Proposed Model}

SNNs process information through discrete spike events rathen than continuous-valued activations. A spiking neuron integrates its inputs in a membrane potential and emits a spike when this potential reaches a threshold. In this work, we employ the leaky integrate-and-fire (LIF) neuron model \cite{Eshraghian2021}, whose spike generation is described by
\begin{equation}
    s_{t,i}^{(l)} = \Theta \left( u_{t,i}^{(l)} - u_\mathrm{th} \right),
\end{equation}
where $u_{t,i}^{(l)}$ is the membrane potential at timestep $t$, for the $i$-th neuron of the $l$-th layer of the SNN, and $\Theta\left(\cdot\right)$ is the Heaviside step function.
The corresponding discrete-time membrane dynamics are
\begin{equation}\label{eq:lifneuron}
    u_{t,i}^{(l)} = \beta u_{t-1,i}^{(l)} + \sum_j{W_{i,j}^{(l)}s_{t-1,j}^{(l-1)}} - u_\mathrm{rst}s_{t,i}^{(l)},
\end{equation}
where $W_{i,j}^{(l)}$ is the weight connecting pre-synaptic neuron $j$ to
post-synaptic neuron $i$, $\beta$ is the membrane-potential decay rate, and
$u_{\mathrm{rst}}$ is the value subtracted from the membrane potential when a spike occurs.

The LIF model can be adapted to obtain the neuron dynamics required for different stages of the proposed SNN. Following \cite{Eshraghian2021}, the decay rate is related to the time constant $\tau$ by
\begin{equation}\label{eq:decayrate}
  \beta = 1 - \frac{1}{\tau}.
\end{equation}

In regression tasks, the output layer typically consists from leaky integrators (LIs) neuron to accumulate incoming spikes without generating output spikes. This is achieved by disabling the reset mechanism and setting the firing threshold sufficiently high to prevent spike generation. The LI output therefore provides a continuous-valued representation, which is appropriate for the positioning regression task \cite{Henkes2024,Huang2025,Respati2026}.

Fully connected SNN architectures require multi-dimensional input tensors to be flattened prior to processing; thereby, destroying their inherent spatial structure. However, tensors derived from CSI measurements exhibit meaningful local spatial correlations between neighboring elements. In CSI data, adjacent tensor elements correspond to nearby antenna elements experiencing similar propagation conditions, while distant URPA sub-sectors are typically less correlated. Consequently, flattening eliminates informative spatial relationships that are beneficial for network convergence and localization accuracy. CSNNs address this limitation by preserving the spatial structure of the input tensors during processing.

CSNNs are constructed similarly to CNNs. They consist of components, namely convolutional and linear components. The convolutional component behaves exactly like its analogous counterpart, with the only difference being that instead of ReLU layers, the CSNN uses LIF layers \cite{Rueckauer2017}. Additionally, batch normalization layers are not required as the internal representation of the CSNN is binary. The output of the ReLU layer and the firing rate of the output spikes of an LIF neuron are proportional to the input. In other words, CSNN can be seen as a modified CNN \cite{Diehl2015}.

Fig.~\ref{fig:architectures} presents the architecture of the proposed CSNN. The input tensor is first transformed into a channel-first format and subsequently processed through 3 convolutional layers integrated with LIF neurons. Each convolutional stage extracts progressively higher-level spatial representations while maintaining temporal membrane dynamics across multiple inference time steps. An adaptive average pooling layer compresses the spatial dimensions into a compact representation, which is then flattened and passed through a spiking regressor head, with a single $N_n$-neuron layer. The output layers is comprised by LI neurons, to effectively produce the coordinate estimates.

\subsection{Computational and Storage Complexity}
\label{subsec:consumption}
Computational complexity is defined by the number of 32-bit floating-point operations (FLOPs), required for a single inference. A FLOP can either be an accumulation (AC) or a multiplication-and-accumulation (MAC). The number of FLOPs is proportional to the size of the model architecture. For a conventional, linear dense network, i.e., a regressor head, with $N_l$ layers, the number of FLOPs required is
\begin{equation}
    O_\mathrm{Dense} = \sum_{l=1}^{N_l}{N_\mathrm{in}^{(l)}N_{\mathrm{out}}^{(l)}}.
\end{equation}
Similarly, for a convolutional segment with $N_l$ convolutional layers, the number of FLOPs is defined as
\begin{equation}\label{eq:cnn_flops}
    O_\mathrm{Conv} = \sum_{l=1}^{N_l}{H_\mathrm{out}^{(l)}W_\mathrm{out}^{(l)}\left(M_\mathrm{K}^{(l)}\right)^2  C_\mathrm{in}^{(l)}C_{\mathrm{out}}^{(l)}},
\end{equation}
where $H_\mathrm{out}^{(l)}$ and $W_\mathrm{out}^{(l)}$ are the height of the input at layer $l$, respectively. Consequently, when an architecture contains convolutional and dense segments, the total number of FLOPs is the summation of the respective number of FLOPs in each segment, i.e., $O_\mathrm{Dense} + O_\mathrm{Conv}$.

For spiking models, their computational complexity is proportional to the firing rate \cite{Kundu2021}; defined as
\begin{equation}
    \zeta^{(l)} = \frac{O_\mathrm{spike}^{(l)}}{O_\mathrm{neuron}^{(l)}},
\end{equation}
where $O_\mathrm{spike}^{(l)}$ is the total number of spikes generated in layer $l$ for all timesteps $T$, and $O_\mathrm{neuron}^{(l)}$ is the total number of neurons in layer $l$. Each spike generation in layer $l$ is followed by an accumulation in layer $l+1$, therefore the number of accumulations, $O_\mathrm{AC}$ is proportional to the number of spikes \cite{Kundu2021}. For an $N_l$-layered fully-connected SNN, the total number of FLOPs is
\begin{equation}\label{eq:snn_flops}
    O_\mathrm{SNN} = \sum_{l=1}^{N_l}{\zeta^{(l)}N_\mathrm{in}^{(l)}N_{\mathrm{out}}^{(l)}}.
\end{equation}
Similarly, the total number of FLOPs for an $N_l$-layered CSNN is
\begin{equation}\label{eq:csnn_flops}
    O_\mathrm{CSNN} = \sum_{l=1}^{N_l}{\zeta^{(l)}H_\mathrm{out}^{(l)}W_\mathrm{out}^{(l)}\left(M_\mathrm{K}^{(l)}\right)^2  C_\mathrm{in}^{(l)}C_{\mathrm{out}}^{(l)}}.
\end{equation}

Although, as seen in (\ref{eq:lifneuron}), a weighted spike mathematically involves the product $W_{i,j}^{(l)}s_{t-1,j}^{(l-1)}$, for binary spikes an event-driven implementation can realize this operation as a conditional accumulation of the corresponding weight. Therefore, the computational cost is commonly modeled in terms of AC, instead of MAC operations.

Storage complexity is defined by the number of parameters, i.e. weights and biases. A model with $N_p$ parameters requires $\mathcal{O}\left(N_p\right)$ memory. For example, in a 32-bit computational system, a model with $N_p$ parameters, requires $4\times N_p$ bytes of memory space.

For the $k$-NN baseline, each query requires an exhaustive distance computation against the entire fingerprint database, so its cost is $O_{k\text{-NN}} = N_{\mathrm{db}}D$ MACs, where $N_{\mathrm{db}}$ is the number of stored fingerprints and $D$ their dimensionality. Since the radio map itself constitutes the model, the storage requirement is $4N_{\mathrm{db}}D$ bytes rather than $4N_p$.

\section{PERFORMANCE METRICS}\label{sec:perf_metrics}

The performance of the proposed neural models is evaluated on the test dataset using metrics that assess both positioning accuracy and the resulting communication performance. Specifically, we consider the absolute positioning error (APE), mean positioning error (MPE), root mean squared error (RMSE), and 90th-percentile circular error (CE90). In addition, the impact of the predicted UE position on the communication system is evaluated in terms of the achievable signal-to-noise ratio (SNR) and EE. The definitions of these metrics are provided in the following subsections.

\subsection{Absolute Positioning Error}

The neural network models return an estimated wireless user position $\hat{\textbf{p}} \in \mathbb{R}^{2}$. The absolute positioning error (APE) is calculated as the Euclidean norm between the estimated and the ground-truth position $\mathbf{p}$
\begin{equation}\label{eq:distance}
  d = || \hat{\mathbf{p}} - \mathbf{p} ||_2 = \sqrt{\left( \hat{x} - x \right)^2 + \left( \hat{y} - y \right)^2}.
\end{equation}

The geometric interpretation of APE is a circle around the true position.  

\subsection{Mean Positionning Error}
The mean positioning error (MPE) over $N_d$ samples is defined as
\begin{equation}\label{eq:mean_error}
    \bar{d} = \frac{1}{N_d}\sum_{i}{d_i}.
\end{equation}
The MPE provides a single measure of the average positioning accuracy over
the evaluated samples.

\subsection{Root Mean Square Error}
The root mean squared error (RMSE) represents the errors in the same measuring units as the original data. Mathematically, RMSE measures the standard deviation of the error and is defined as:
\begin{equation}\label{eq:rmse}
  \mathrm{RMSE} = \sqrt{\mathrm{MSE}} = \sqrt{\frac{1}{N_d}\sum_{i=1}^{N_d}{{d_i}^2}}.
\end{equation}

\subsection{$90$\textsuperscript{th} Percentile Circular Error}

Mean-based metrics such as the MPE and the RMSE summarize the central tendency of the error distribution, but they are insensitive to its upper tail \cite{Zandbergen2008}. In beam management the tail is the quantity of operational interest; a small fraction of large positioning errors produces beam misalignment and a corresponding loss of received power, whereas the average error over all users does not by itself characterize link reliability. Percentile-based circular error metrics are therefore the accepted way of specifying positioning accuracy in both geospatial standards and cellular positioning requirements.

Let $d$ denote the APE defined in (\ref{eq:distance}), regarded as a non-negative random variable with CDF
\begin{equation}
    F_d(x) = P\left(d<x\right),~x \geq 0,
\end{equation}
where $x$ is an arbitrary reference distance.

The circular error (CE) at probability level $p$, denoted $\mathrm{CE}_p$, is the $p$-th quantile of $F_d$,
\begin{equation}
    \mathrm{CE}_p = {F_d}^{-1}(p) = \mathrm{inf}\{x \geq 0: F_d(x) \geq p\},
\end{equation}
that is, the radius of the circle centered on the true position that contains a fraction $p$ of the position estimates. Setting $p=0.9$ yields the $90^{\mathrm{th}}$ percentile circular error,
\begin{equation}
    P\left(d \leq \mathrm{CE}_{90}\right) = 0.9.
\end{equation}

$F_d$ is not available in closed form for a learned estimator, so $\mathrm{CE}_p$ is obtained from the empirical distribution of the $N_d$ evaluated samples.

\subsection{Telecommunications Metrics}
To check the effectiveness of the system model, we use the average SNR and the EE as telecommunication metrics. The downlink beam is steered towards the estimated position $\hat{\mathbf{p}}$ by the analog phase-shifter network defined in~(\ref{eq:FDL}). For a single served UE with all phase shifters active, the precoding vector $\mathbf{w}(\hat{\mathbf{p}})\in\mathbb{C}^{N_{TX}}$ has entries
\begin{equation}\label{eq:precoder}
    w_i(\hat{\mathbf{p}}) = \frac{1}{\sqrt{N_{TX}}}\,
    e^{j\theta_{S_i}}, \qquad
    \theta_{S_i} = \frac{2\pi}{\lambda_c}\,
    \mathbf{u}_i^T\hat{\mathbf{k}}(\hat{\mathbf{p}}),
\end{equation}
where $\mathbf{u}_i\in\mathbb{R}^3$ is the position of the $i$-th array element relative to the array phase centre and
\begin{equation}\label{eq:steervec}
    \hat{\mathbf{k}}(\hat{\mathbf{p}}) =
    \frac{\hat{\mathbf{p}}-\mathbf{p}_{BS}}
         {\left\lVert \hat{\mathbf{p}}-\mathbf{p}_{BS} \right\rVert_2}
\end{equation}
is the unit vector from the BS position towards the estimated UE position. The effective single-input channel observed by the UE is then the beamformed combination
\begin{equation}\label{eq:beamformed}
    \tilde{h}_k\!\left(\hat{\mathbf{p}}\right) =
    \mathbf{w}^H(\hat{\mathbf{p}})\,\mathbf{h}_k,
\end{equation}
with $\mathbf{h}_k\in\mathbb{C}^{N_{TX}}$ denoting the per-element CFR of~(\ref{CFR}) evaluated per subcarrier. The achievable SNR at the estimated position follows as
\begin{equation}\label{eq:snr}
    \mathrm{SNR}_{\mathrm{dB}}(\hat{\mathbf{p}}) =
    10\log_{10}\!\left(\frac{1}{N}\sum_{n=1}^{N}
    \left| \tilde{h}_k\!\left(\hat{\mathbf{p}}\right)\right|^2\right)
    - P_{\mathrm{noise}},
\end{equation}
where $P_{\mathrm{noise}}$ is the receiver noise power. Setting $\hat{\mathbf{p}}=\mathbf{p}$ in~(\ref{eq:precoder}) yields the genie-aided reference
$\mathrm{SNR_{genie}} = \mathrm{SNR}_{\mathrm{dB}}(\mathbf{p})$, against which every model is compared. 

Additionally, we define the link-level EE, as the spectral efficiency (SE) of the downlink divided by the total power consumed to deliver it.
\begin{equation}
    \mathrm{EE}=\frac{\mathrm{SE}}{P_\mathrm{tot}},
\end{equation}
with $\rm{SE}$ and $E_\mathrm{tot}$ defined as
\begin{equation}
    \rm{SE}=W \, \log_2\left(1 + 10^{\frac{\mathrm{SNR}_{\mathrm{dB}}(\hat{\mathbf{p}})}{10}}\right),
\end{equation}\begin{equation}
    E_\mathrm{tot} = \frac{P_{t}}{\mathrm{PA}_e} + P_\mathrm{co} + K_\mathrm{dl}\,P_\mathrm{chain} + N_\mathrm{TX}\,P_\mathrm{PS} + P_\mathrm{inf},
\end{equation}
respectively.

The $W$ represents the operational bandwidth, $P_{t}$ the transmission power, $\mathrm{PA}_e$ the  power amplifier efficiency of the transmitter, $P_\mathrm{co}$ the baseline control power of the circuitry, $P_\mathrm{chain}$ and $P_\mathrm{PS}$ the RF chain power and phase shifter power consumption, respectively. Finally, $P_\mathrm{inf}$ represents the power consumption introduced by the position estimator. For a single user updating once, $E_\mathrm{inf}$  is of order $10^{-4} J$~ against a link power of tens of watts. In this scenario, the saving is negligible. The energy efficiency benefits become material only when the positioning subsystem runs continuously for many users. Therefore, $P_\mathrm{inf}$ scales with the number of served UES, $K$, and the beam-update interval time, $\Delta t$, as
\begin{equation}
    P_\mathrm{inf} = \frac{KE_\mathrm{inf}}{\Delta t}.
\end{equation}

\section{SIMULATION RESULTS \& DISCUSSIONS}\label{sec:results}

\subsection{Simulation Scenario}

Ray-tracing channels are generated according to (\ref{CFR}) using the NVIDIA Sionna-RT$^{\copyright}$ simulator \cite{sionna2022}. In the BS, we assume $N_{BS}=1056$, $N_{RX}=32$ with $N^x_{RX}=8$ and $N^y_{RX}=4$, $N_{TX}=1024$ and $N_{RF}=2$ RF chains. The OFDM signal comprises $N=1024$ subcarriers, $f_c=30\,\mathrm{GHz}$ and $\Delta f=2^{\mu}\cdot15\,\mathrm{kHz}$ with $\mu\in\mathbb{N^*}$ representing the OFDM numerology parameter and $\Delta f\geq 0.9/\tau_{_{DS}}$ where $\tau_{_{DS}}$ represents the RMS of the delay spread, whether $\sigma_{_{RT}}=0.0032\,\mathrm{dBm}$ and $\mu_0=0$. In terms of telecommunication metrics, we denote $N_{av}=4$, $W=122.88\,\mathrm{MHz}$, $P_\mathrm{noise}=-114\,\mathrm{dB}$, $P_{t}=1\,\mathrm{W}$, $\mathrm{PA}_e=0.3$, $P_\mathrm{co}=10\,\mathrm{W}$, $P_\mathrm{chain}=0.250\,\mathrm{W}$ and $P_\mathrm{PS}=0.030\,\mathrm{W}$.

Three datasets are generated, each containing $30,800$ samples. The datasets differ by a positioning offset of $0.08~\mathrm{m}$. The number of samples $B$ corresponds to the number of user positions on the considered grid. The simulation is not inherently a time-series problem. Instead, the user moves across discrete positions on the grid, stops at each position, and transmits a pilot OFDM sequence towards the BS's receiver. Subsequently, the BS receives the signal and estimates the CSI using least-squares (LS) channel estimation for OFDM MIMO systems. The estimated CSI values are collected and stored into tensors. Each CSI tensor has dimensions $[B,~32,~32]$. At the end of the final preprocessing step, each CSI tensor has dimensions $[B,~32,~32,~2]$. The channel first illustration of a tensor can be seen in Fig.~\ref{fig:architectures}. 

The evaluation subset is offset from the training grid by only $0.08~\mathrm{m}$, so every query lies close to a stored fingerprint; a nearest-neighbor estimator therefore operates under favorable conditions. The number of neighbors was swept over $k \in \{ 1,3,5,7,9,15 \}$ on a held-out partition of the fingerprint database, and the reported configuration $k=9$ is the best of the sweep.

All neural models were trained on $61,600$ samples using hyperparameters selected through empirical experimentation. The antenna is positioned at $(133,~204)$. Therefore the dataset translation is applied for $dx=133$ and $dy=204$.

The neural network models were trained using model-specific batch sizes, and numbers of training epochs. Training was performed sequentially with multiple batch sizes to capture patterns at different levels of granularity \cite{smith2018dontdecaylearningrate, dichasus2021}. The CNN model was trained for $250$ epochs with increasing batch size every $50$ epochs $\{32,~64,~256,~1024$,~$4096\}$. The CSNN model was trained for $400$ epochs, with increasing batch size every $100$ epochs, drawn from the set $\{32,~64,~256,~1024\}$.

All models were trained using the MSE loss function, defined as
\begin{equation}\label{eq:mse_loss}
\mathcal{L} = \frac{1}{2B}\sum_{i=1}^{B}{\left[ \left( \hat{x_i} - x_i \right)^2 + \left( \hat{y_i} - y_i \right)^2 \right]}.
\end{equation}

For the brain-inspired models, the surrogate gradients chosen were; the default, $\arctan(\pi u_t)$, and the fast sigmoid $u_t/\left(1+ku_t\right)$, respectively. In the fast sigmoid, $k$ is defined as the slope and defaults to $25$. As for the reset mechanism, we opted for a `zero' reset, where $u_\mathrm{rst} = \beta u_{t-1,i}^{(l)} + \sum_j{W_{i,j}^{(l)}s_{t-1,j}^{(l-1)}}$, thus driving the membrane potential back to $0$ upon firing a spike.

The number of neurons in the regressor head is $256$ in both the CNN and CSNN. Both convolutional models define $C_\mathrm{in}=32$, $M_\mathrm{K}=3$, and $M_\mathrm{F}=4$. The CSNN was trained three times with $T \in \{5,~10,~15\}$.  The learning rate was kept at its default value of $10^{-3}$ for both models. Furthermore, both models have $N_\mathrm{out}=2$. The models were additionally validated on previously unseen data extracted from the same simulation environment as the training dataset.

\begin{figure}[t]
    \centering
    \subfloat[9-NN model\label{fig:9NN_dist}]{\includegraphics[width=\columnwidth]{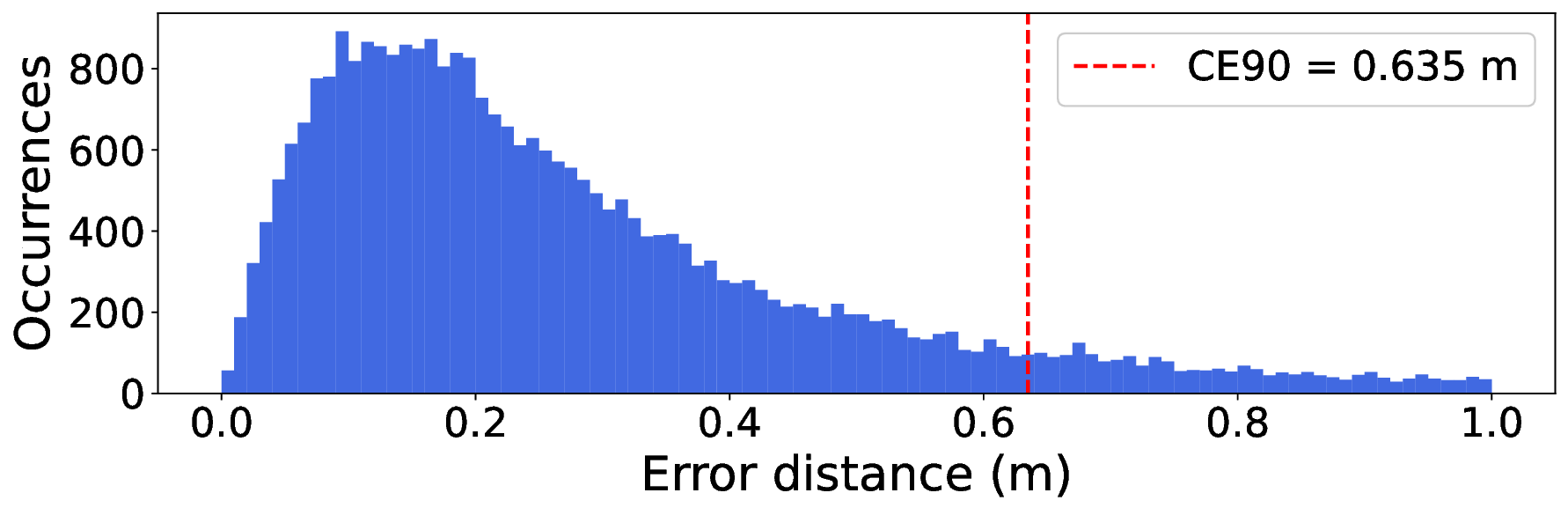}}\\
    \subfloat[CNN model\label{fig:cnn_dist}]{\includegraphics[width=\columnwidth]{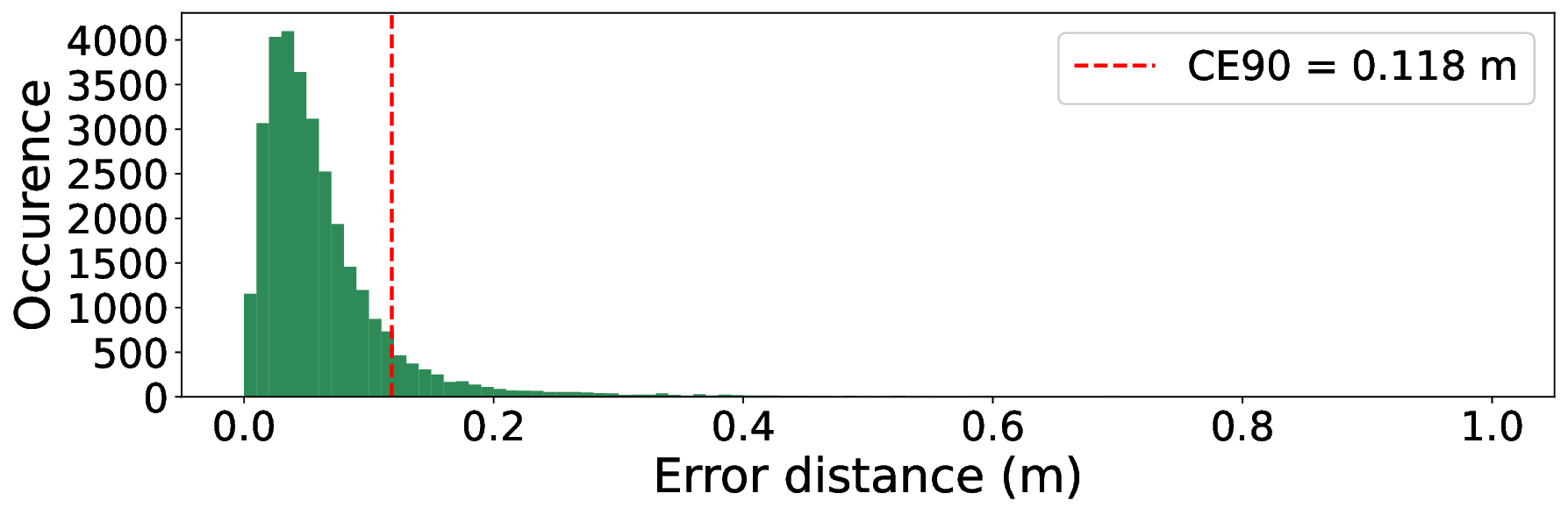}}\\
    \subfloat[CSNN (T = 20) model\label{fig:csnn_dist}]{\includegraphics[width=\columnwidth]{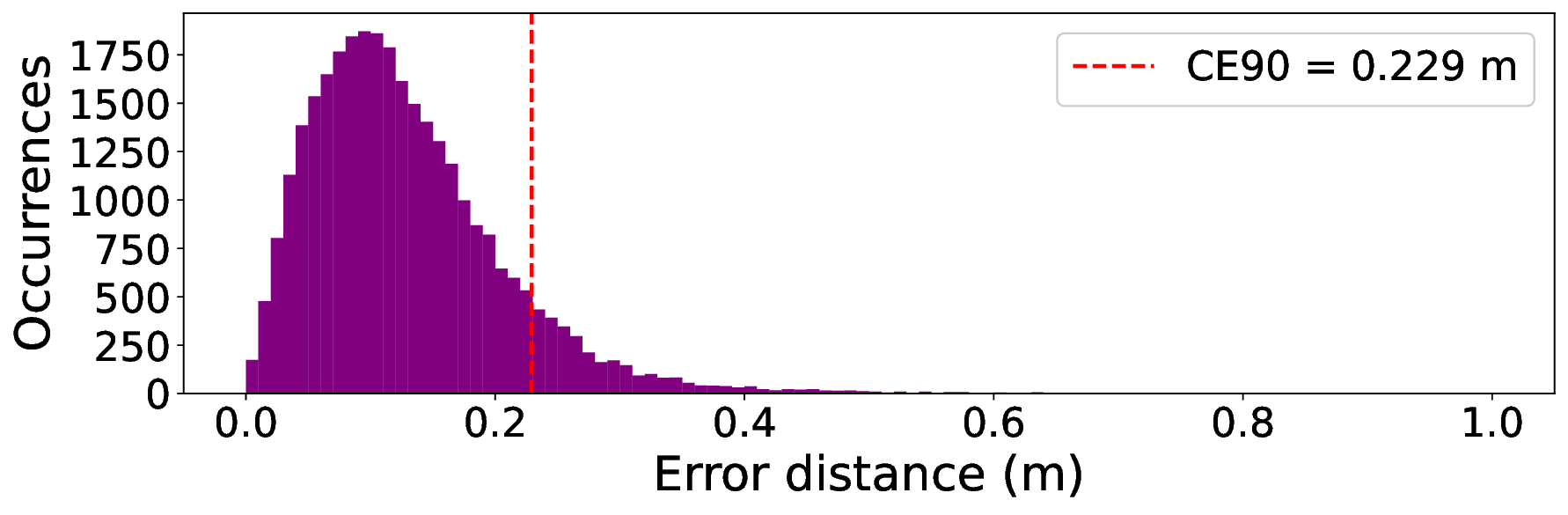}}
    \caption{Distribution of errors for the ANN, CNN, and CSNN (T=20) models.}
    \label{fig:error_distributions}
\end{figure}

\subsection{Positioning Error Distribution}

The models are first assessed through the distribution of the absolute positioning error defined in~(\ref{eq:distance}). Fig.~\ref{fig:error_distributions} depicts the error histograms of the three schemes, and Table~\ref{tab:positioning} summarizes the corresponding percentile and moment-based statistics.

The CNN attains the lowest error of the three models, with $\mathrm{CE}_{90}=0.118~\mathrm{m}$. The proposed CSNN follows at $\mathrm{CE}_{90} = 0.229~\mathrm{m}$, i.e. a factor of $1.9$ higher, while the 9-NN fingerprinting baseline reaches $\mathrm{CE}_{90}=0.635~\mathrm{m}$. The ordering is preserved across every reported statistic, so no scheme is favored by the choice of metric.

The 9-NN result merits particular attention, because it bounds what can be achieved by interpolation over the radio map alone. Even so, 9-NN remains $5.4$ times less accurate than the CNN and $2.8$ times less accurate than the CSNN. The learned models therefore extract structure from the CSI that weighted averaging neighboring fingerprints does not recover, which is consistent with the wavelength scale at $30~\mathrm{GHz}$: with $\lambda_c=1~\mathrm{cm}$, the channel de-correlates over distances far shorter than the grid spacing, so spatially adjacent fingerprints are not necessarily similar in the CSI domain.

Comparing the two learned models, the accuracy gap between CNN and the CSNN is the direct cost of the spiking representation. Both share an identical layer structure, differing only in that the ReLU activations of the CNN are replaced by LIF neurons and that the network is unrolled over $T$ time steps. The internal representation of the CSNN is consequently binary, and the information carried by a real-valued activation must instead be encoded in the number and timing of spikes emitted within the available time steps. The resulting loss of representational precision accounts for the increase in positioning error, and is quantified as a function of $T$ in Section~\ref{subsec:timesteps}.

\begin{figure*}[ht]
    \centering
    \subfloat[k-NN model\label{fig:area}]
    {\includegraphics[width=0.33\textwidth]{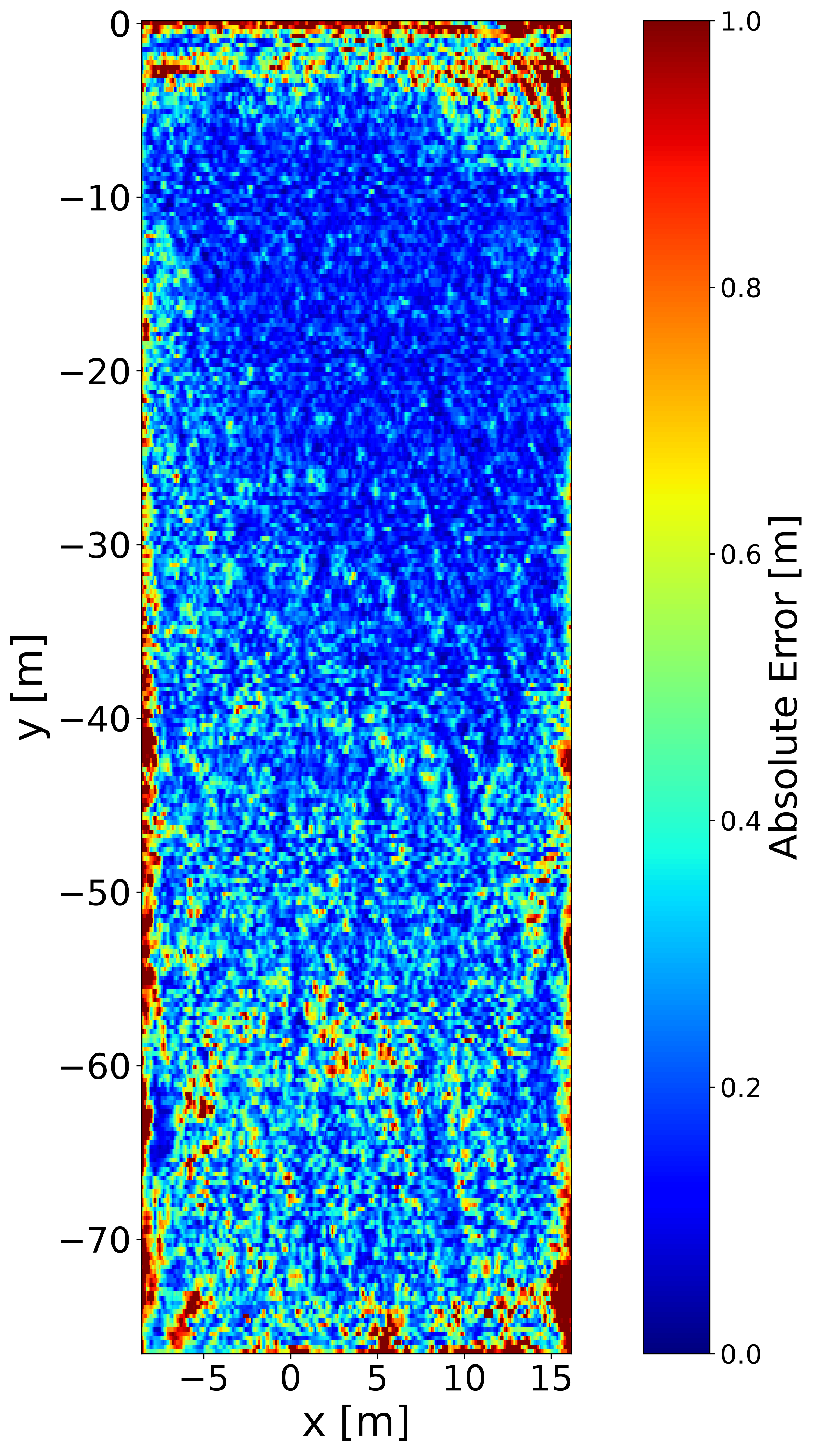}}
    \hfill
    \subfloat[CNN model\label{fig:cnn_area}]
    {\includegraphics[width=0.33\textwidth]{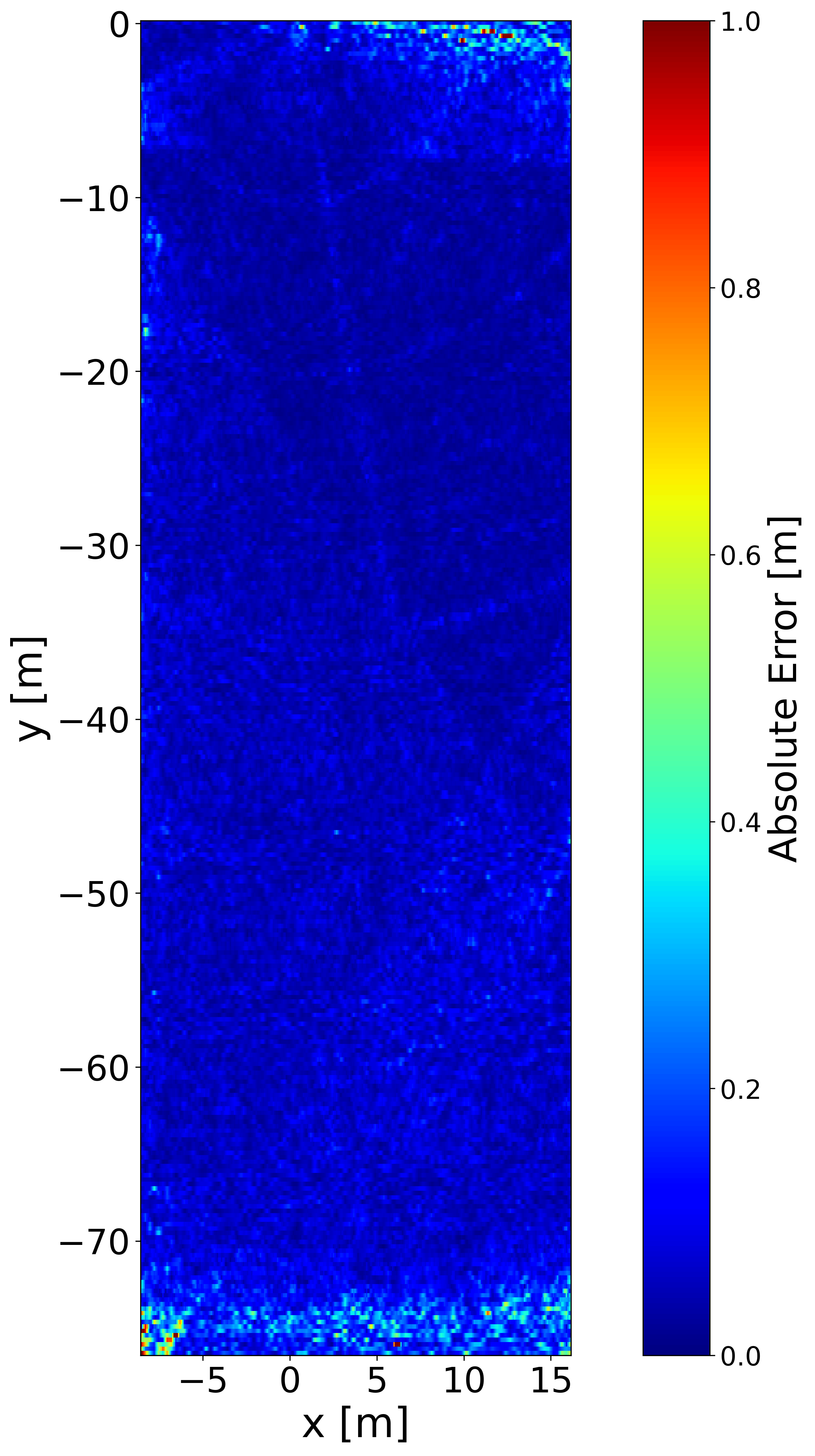}}
    \hfill
    \subfloat[CSNN (T=20) model\label{fig:csnn_area}]
    {\includegraphics[width=0.33\textwidth]{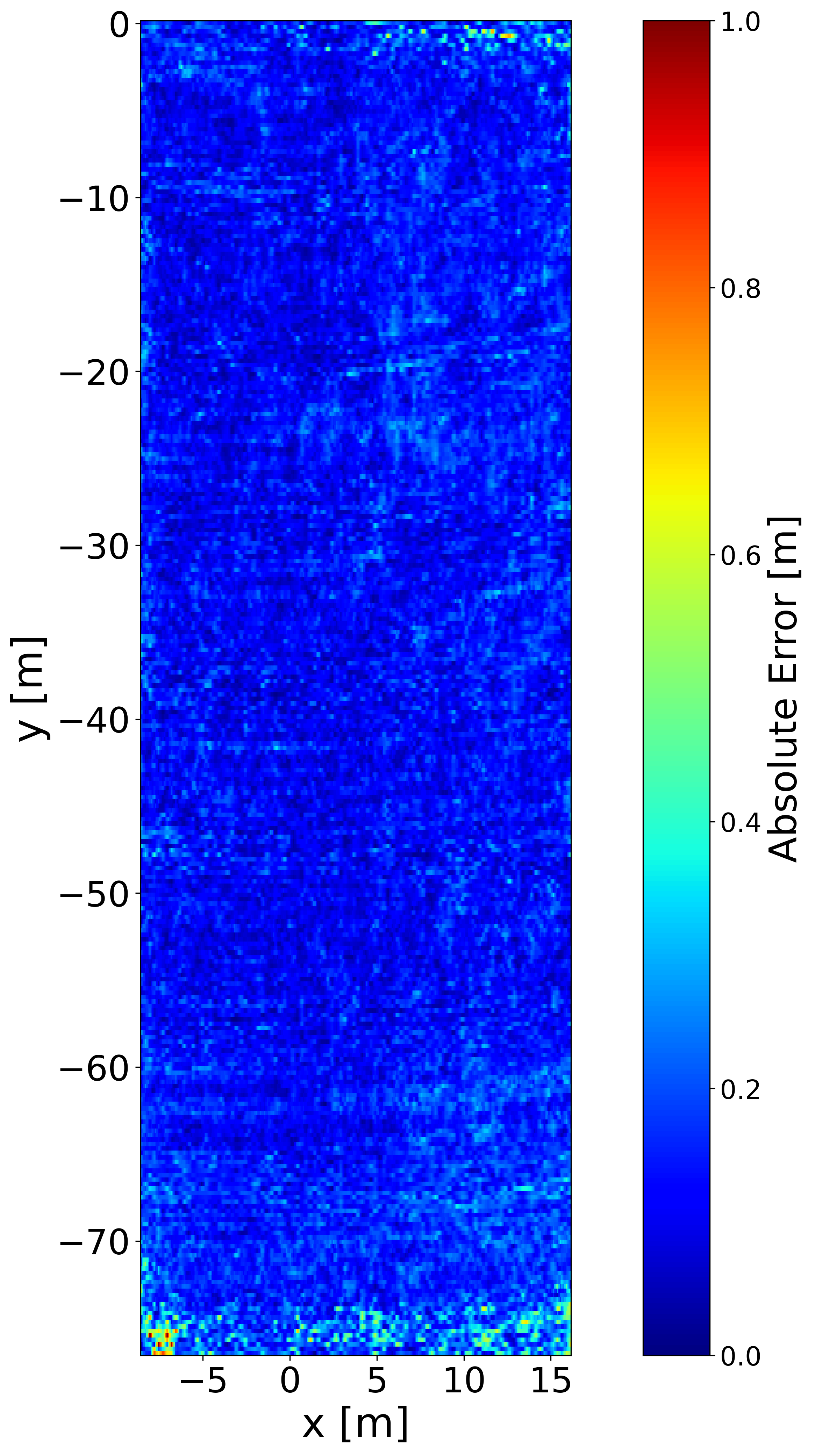}}
    \hfill
    \caption{Distribution of APEs in space for the ANN, CNN, and CSNN (T=20).}
    \label{fig:spatial_error_distribution}
\end{figure*}

The spatial distribution of the error is illustrated in Fig.~\ref{fig:spatial_error_distribution}, obtained by interpolating the per-position errors onto a dense grid. All three schemes eschibit their largest errors in the region closest to the BS and near the boundaries of the surveyed area. 

\begin{table}[t]
\centering
\caption{Positioning accuracy of the evaluated schemes on the held-out
evaluation subset. All values in metres.}
\label{tab:positioning}
\renewcommand{\arraystretch}{1.2}
\begin{tabular*}{\columnwidth}{@{\extracolsep{\fill}}lccccc}
\toprule
Scheme & CE50 & CE90 & CE95 & MPE & RMSE \\
\midrule
CNN            & 0.048 & \textbf{0.118} & 0.162
               & 0.064 & 0.094 \\
CSNN ($T=20$)  & 0.115 & 0.229 & 0.271
               & 0.130 & 0.154 \\
CSNN ($T=10$)  & 0.135 & 0.277 & 0.334
               & 0.154 & 0.182 \\
CSNN ($T=5$)   & 0.123 & 0.244 & 0.295
               & 0.139 & 0.166 \\
$9$-NN         & 0.225 & 0.635 & 0.836 & 0.311 & 0.646 \\
\bottomrule
\end{tabular*}
\end{table}

\begin{figure}[t]
    \centering
    \includegraphics[width=\columnwidth]{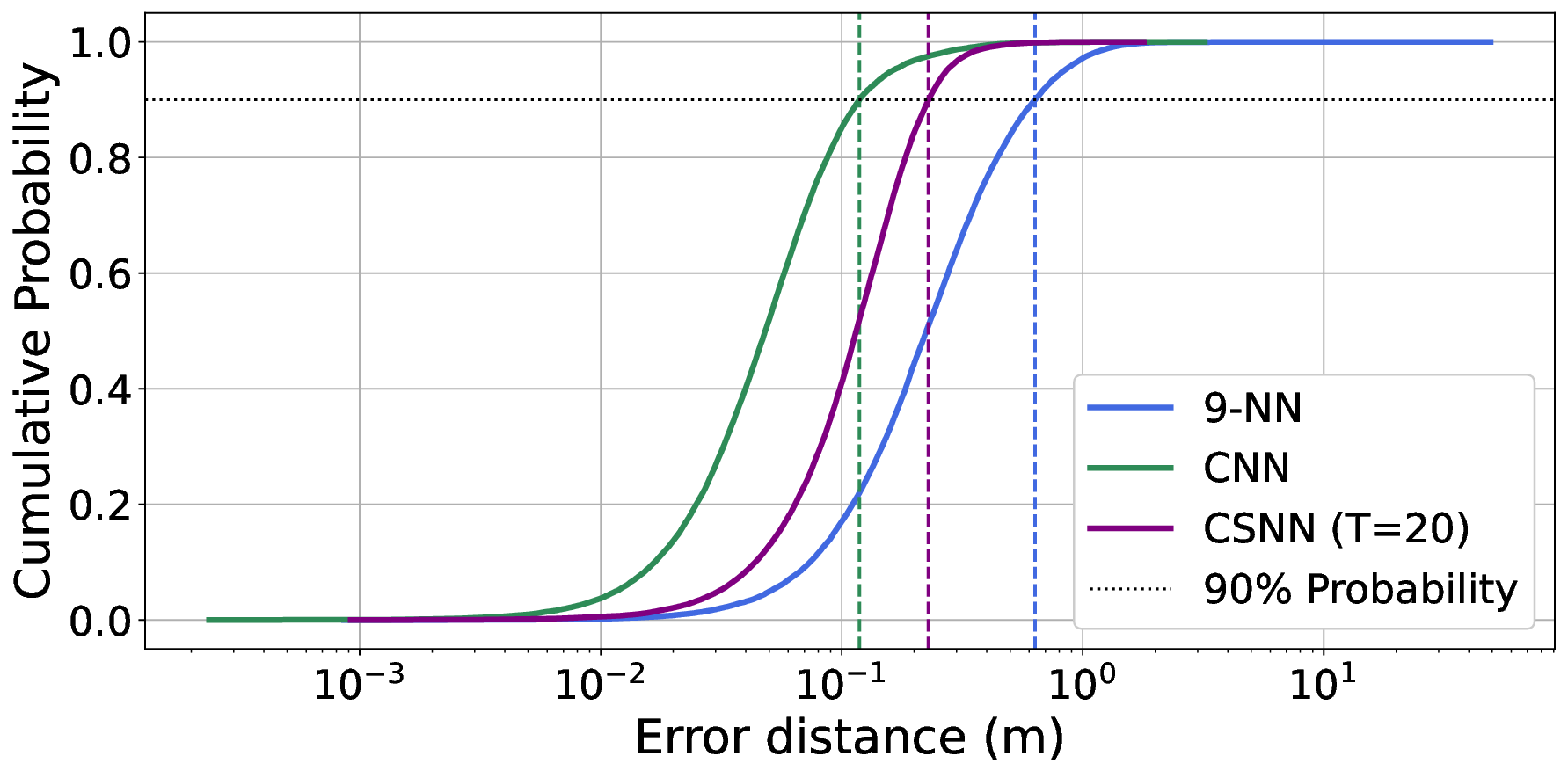}
    \caption{Positioning error CDF curves.}
    \label{fig:CDFs}
\end{figure}

\subsection{Effect of Number of Time Steps}\label{subsec:timesteps}

The number of time steps $T$ governs both the temporal resolution available to the LIF dynamics, and the number of synaptic operations executed per inference. To characterise this trade-off, the CSNN was trained at $T \in \{5, 10, 20\}$ under an otherwise identical configuration, and the per-layer firing rates were measured over the evaluation subset.

\begin{table}[t]
\centering
\caption{Effect of the number of time steps $T$ on CSNN positioning accuracy,
spiking activity and estimated inference energy.}
\label{tab:timesteps}
\renewcommand{\arraystretch}{1.2}
\begin{tabular*}{\columnwidth}{@{\extracolsep{\fill}}lccc}
\toprule
 & $T=5$ & $T=10$ & $T=20$ \\
\midrule
CE90 [m]                                    & 0.2443 & 0.2767 & \textbf{0.2293} \\
Mean firing rate $\bar{r}$                  & 0.0722 & 0.0373 & 0.0398 \\
Spikes per neuron $\zeta=\bar{r}T$          & 0.3609 & 0.3729 & 0.7957 \\
Energy per inference [mJ]                   & \textbf{0.0475} & 0.0525 & 0.1122 \\
\bottomrule
\end{tabular*}
\end{table}

Table~\ref{tab:timesteps} reports the results. Positioning accuracy does not improve monotonically with $T$, with $\mathrm{CE}_{90}=0.244$~m at $T=5$ to $0.277$~m at $T=10$, and down to $\mathrm{CE}_{90}=0.229$~m at $T=20$. Decreasing by $13.5\%$ from $T=5$ to $T=10$, and increasing by $17.3\%$ from $T=10$ to $T=20$. The ordering is not consistent with a simple dependence on temporal resolution; we therefore attribute part of it to run-to-run variation between independently trained networks rather than to $T$ alone.

The energy, by contrast, varies monotonically and more strongly, growing from $0.0475$~mJ at $T=5$ to $0.1122$~mJ at $T=20$. The growth is nevertheless sub-linear in $T$, and the firing statistics of Table~\ref{tab:firing_rates} explain why. Between $T=5$ and $T=10$ the mean per-time-step firing rate $\bar{r}$ falls from $0.072$ to $0.037$, approximately $1/T$, so that the spike count per neuron per inference $\zeta=\bar{r}T$ is essentially unchanged ($0.361$ against $0.373$). Over this range the trained network converges to an approximately fixed spike budget, and $T$ determines only how finely that budget is distributed in time. Since the synaptic operation count in~(\ref{eq:csnn_flops}) scales with $\zeta$ rather than with $T$ directly, doubling the number of time steps increases the energy by only $11\%$.

This compensation does not persist at $T=20$, where $\bar{r}$ rises slightly to $0.040$ and $\zeta$ consequently doubles to $0.796$. The additional time steps are no longer absorbed by a proportional reduction in firing rate, and the energy grows accordingly. The effect is concentrated in two layers. The first convolutional layer, which receives the constant input current at every time step, exhibits $\zeta^{(1)}$ increasing form $1.088$ at $T=5$ to $3.346$ at $T=20$. The spiking regressor head reaches $\zeta^{(4)}=4.432$, approaching the break-even value $E_{\mathrm{MAC}}/E_{\mathrm{AC}}\approx5.11$ above which an accumulation-based layer would cease to be advantageous over its multiply--accumulate counterpart. The two intermediate convolutional layers remain sparse throughout, with $\zeta^{(2)}\leq0.774$ and $\zeta^{(3)}\leq0.163$.

Taken together, these results indicate that the accuracy attainable by the CSNN is only weakly dependent on $T$ over the swept range, whereas the inference energy is not. We adopt $T=20$ in the remainder of this section because it achieves the best measured accuracy, and note that $T=5$ offers a $42\%$ reduction in inference energy for a $6.5\%$ increase in $\mathrm{CE}_{90}$, which is the configuration a deployment constrained by energy rather than accuracy would select.

\begin{table}[t]
\centering
\caption{Per-layer spiking activity of the CSNN measured over the evaluation subset. $\bar{r}^{(l)}$ is the fraction of neurons firing per time step and $\zeta^{(l)}=\bar{r}^{(l)}T$ the number of spikes per neuron per inference. Values are reported with three decimals precision.}
\label{tab:firing_rates}
\renewcommand{\arraystretch}{1.2}
\begin{tabular*}{\columnwidth}{@{\extracolsep{\fill}}lcccccc}
\toprule
& \multicolumn{2}{c}{$T=5$} & \multicolumn{2}{c}{$T=10$}
& \multicolumn{2}{c}{$T=20$} \\
\cmidrule(lr){2-3}\cmidrule(lr){4-5}\cmidrule(lr){6-7}
Layer & $\bar{r}^{(l)}$ & $\zeta^{(l)}$ & $\bar{r}^{(l)}$ & $\zeta^{(l)}$
      & $\bar{r}^{(l)}$ & $\zeta^{(l)}$ \\
\midrule
LIF 1 (conv) & 0.218 & 1.088 & 0.127 & 1.266 & 0.167 & 3.346 \\
LIF 2 (conv) & 0.077 & 0.385 & 0.042 & 0.415 & 0.039 & 0.774 \\
LIF 3 (conv) & 0.033 & 0.163 & 0.012 & 0.125 & 0.008 & 0.162 \\
LIF 4 (FC)   & 0.477 & 2.385 & 0.253 & 2.530 & 0.222 & 4.432 \\
\midrule
Overall      & 0.072 & 0.361 & 0.0373 & 0.373 & 0.040 & 0.796 \\
\bottomrule
\end{tabular*}
\end{table}

\subsection{Energy Consumption}

\begin{figure}[t]
  \centering
\includegraphics[width=\columnwidth]{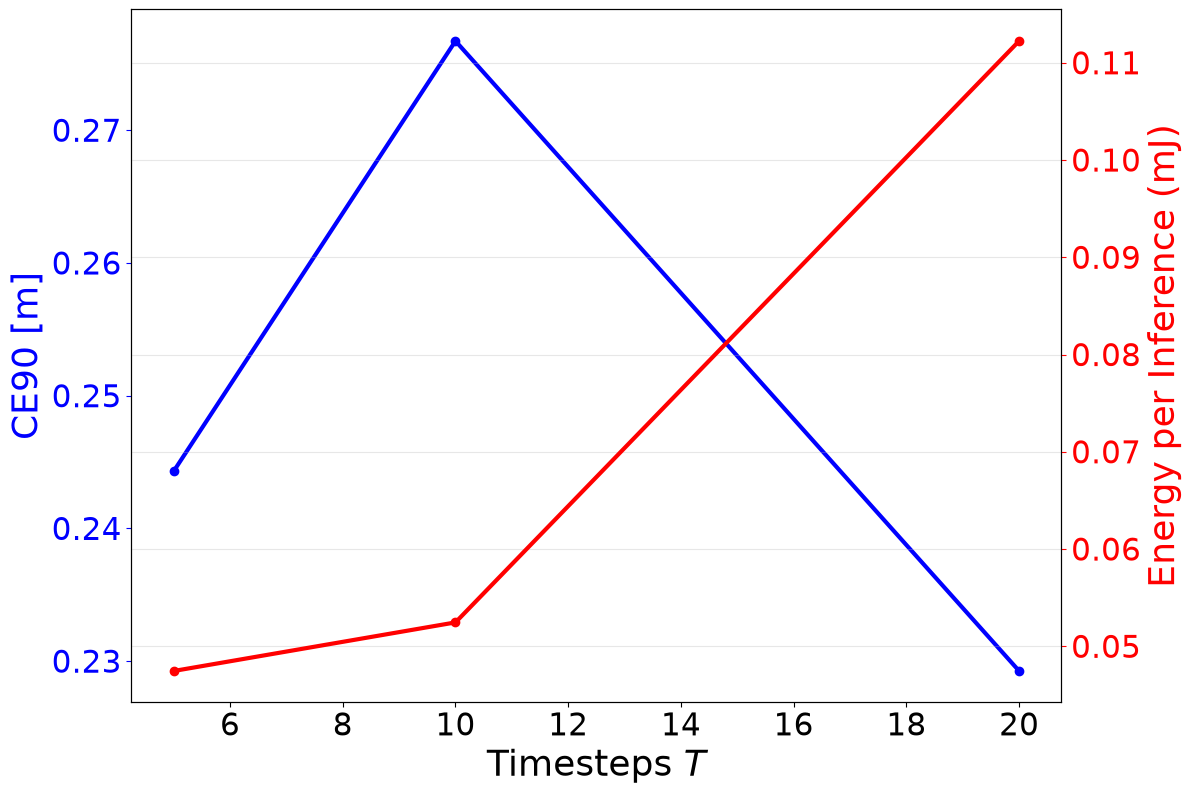}
  \caption{Positioning accuracy (CE90, left axis) and estimated energy per inference (right axis) of the proposed CSNN as a function of the number of time steps $T$. Energy is computed analytically from the measured per-layer firing rates at $45$~nm, and includes all $T$ time steps of a single position estimate.}
  \label{fig:timesteps_tradeoff}
\end{figure}

The preceding results establish the positioning accuracy attainable by each model. Accuracy alone, however, does not determine suitability for deployment at a neuromorphic edge unit, where the inference cost per position estimate is the binding constraint. We therefore turn to the computational energy, first examining how it varies with the number of time steps of the proposed CSNN, and then comparing all models under a common analytical model.

Fig.~\ref{fig:timesteps_tradeoff} shows the CE90 and the estimated energy per inference of the CSNN as a function of $T$. The two quantities are not in simple opposition. The energy grows monotonically with $T$, attaining its maximum at $T=20$, whereas the accuracy varies weakly and non-monotonically and is best at that same value. The origin of the sub-linear energy growth is the spike budget reported in Table~\ref{tab:firing_rates}: the synaptic operation count is governed by $\zeta$ rather than by $T$ directly, and $\zeta$ is approximately preserved between $T=5$ and $T=10$ before doubling at $T=20$. The number of time steps therefore trades accuracy against energy far less steeply than the factor $T$ would suggest.

The computational energy consumption can be measured according to the number of ACs and MACs. For computational chips designed in $45~\mathrm{nm}$ technology, the energy consumption of ACs and MACs is specified to be $0.9~\mathrm{pJ}$ and $4.6~\mathrm{pJ}$ respectively \cite{Horowitz2014}. Hence, the consumption of a spiking model can be estimated by measuring spiking activity \cite{Chen2023Spike}. Since our spiking models use continuous inputs, their input layer outputs are counted as MACs. The remaining layers exchange exclusively spikes, hence their outputs are considered ACs. 

We report analytical estimates for all models rather than measured GPU energy. Per-inference energy measured on a general-purpose GPU is dominated by static power and framework overhead at unit batch size, and is therefore not a property of the model; moreover, a comparison in which conventional models are measured on a GPU while spiking models are evaluated under a neuromorphic operation model is not meaningful. Applying~(\ref{eq:cnn_flops}) and~(\ref{eq:csnn_flops}) with the measured firing rates of Table~\ref{tab:firing_rates} places every model on a common basis.

The CNN executes $95.49$~M MAC operations per inference, corresponding to $0.4392~\mathrm{mJ}$, of which the convolutional stages account for $99.4\%$. The proposed CSNN performs $0.59$~M MAC operations in its input layer, which is evaluated once because the input is constant across time steps, together with $121.69$~M accumulations distributed over the $T=20$ time steps, giving $0.1122~\mathrm{mJ}$. The CSNN therefore requires $3.9\times$ less energy than the CNN of identical layer structure, at a positioning accuracy cost of $\mathrm{CE}_{90}=0.229~\mathrm{m}$ against $0.118~\mathrm{m}$.

This reduction is bounded by the ratio $E_{\mathrm{MAC}}/E_{\mathrm{AC}} \approx 5.11$: a spiking layer consumes less energy than its multiply--accumulate counterpart only while $\zeta^{(l)}<5.11$. The measured values in Table~\ref{tab:firing_rates} satisfy this condition in every layer at every swept $T$, with an aggregate $\zeta=0.796$, so the CSNN operates well inside the favourable regime. The margin is smallest at the spiking regressor head preceding the readout, for which $\zeta^{(4)}=4.432$, and at the first convolutional layer, for which $\zeta^{(1)}=3.346$. Both approach the break-even value, and the readout stage in particular is the component that would limit any further increase in $T$.

For reference, the $k$-NN fingerprinting baseline requires $126.2$~M operations per query and $504.6$~MB of stored fingerprints, since the radio map itself constitutes the model. The energy per query was estimated at $0.5803$~mJ. The proposed model therefore requires $5.2$ times less energy and $200$ times less storage, while being $2.8$ times more accurate.

We note that these figures account for arithmetic operations only. Memory access, which at $45~\mathrm{nm}$ costs approximately $5~\mathrm{pJ}$ per $32$-bit cache read, is excluded. For the conventional models this contributes of order $1\%$ of the arithmetic energy, whereas for the spiking models the membrane state of every neuron must be read and written at each time step, scaling as $O(N_{\mathrm{neurons}}T)$ and reducing the reported advantage by roughfly a factor of two under a conservative cache assumption. On dedicated neuromorphic hardware, where membrane state is co-located with the computing element, this penalty is substantially smaller. The reported values should accordingly be interpreted as an arithmetic lower bound.

\subsection{Performance Evaluation of the Communication Link}

To assess the impact of positioning errors on the communication performance, we evaluate the signal-to-noise ratio (SNR) and energy efficiency (EE) of the downlink transmission. For each UE position, the proposed positioning models are used to estimate the UE location at the BS. The estimated position is then used to determine the downlink transmission beamforming direction. The resulting SNR and EE are compared with the corresponding values obtained under perfect knowledge of the UE position.

For a given neural network model $\mathcal{X}$, where $\mathcal{X} \in
    \{\mathrm{C},\mathrm{CS},\mathrm{K}\}$, the symbols $\mathrm{C}$, $\mathrm{CS}$, and $\mathrm{K}$ denote the CNN, CSNN, and $k$-NN models, respectively.

The absolute SNR error is defined as
$|\Delta \mathrm{SNR}|= |\mathrm{SNR_{genie}} - \mathrm{SNR}_\mathcal{X}|$,
where $\mathrm{SNR_{genie}}$ denotes the SNR obtained when the BS has perfect knowledge of the UE position, while $\mathrm{SNR}_\mathcal{X}$ denotes the SNR obtained when the UE position is estimated using model $\mathcal{X}$. 

\begin{figure}[t]
    \centering
    \includegraphics[width=\columnwidth]{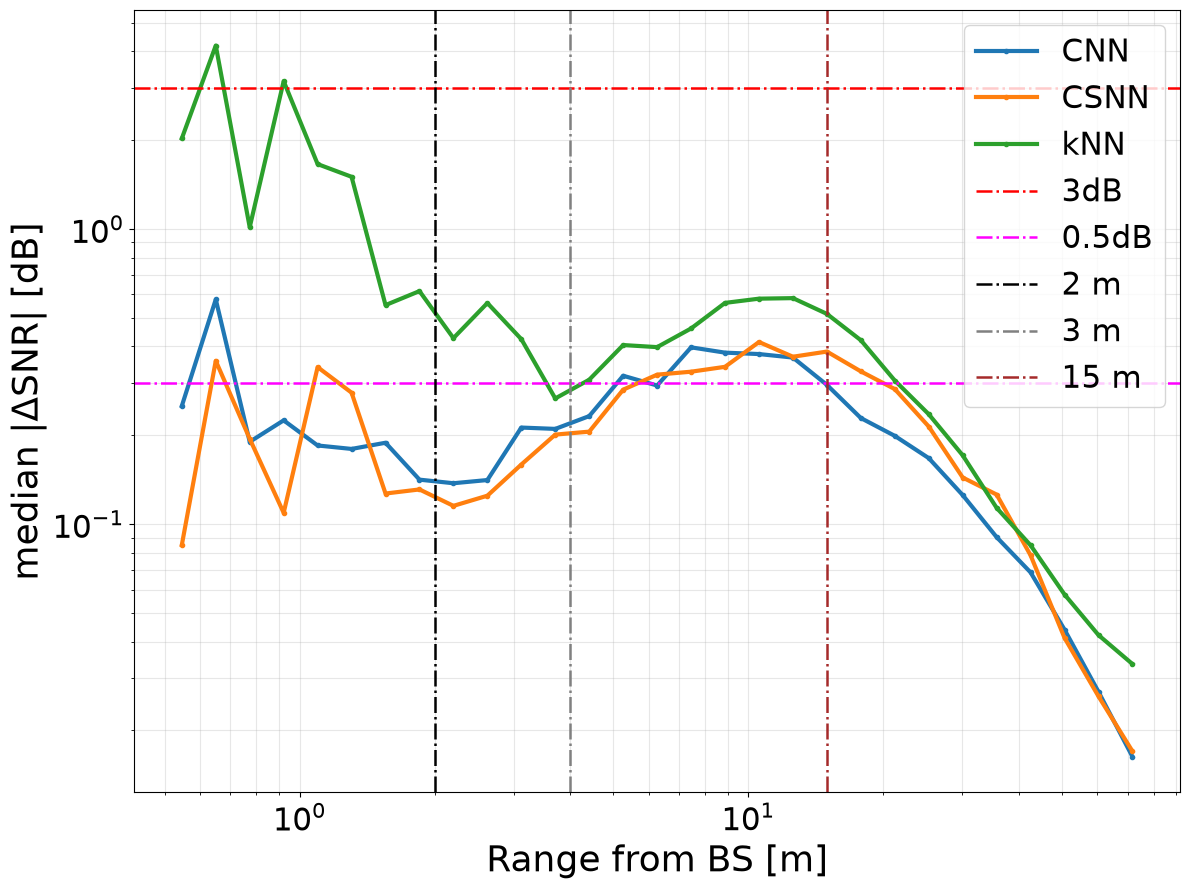}
    \caption{Effect of beam misalignment on optimal receive SNR. The figure measures the difference bewtween the genie-aided beamforming and the beamforming based on each model. Significant degradation is observed near the antenna, before the 3 meters mark. Degradation becomes insignificant the further away the receiver is.}
    \label{fig:BMP}
\end{figure}

Similarly, the absolute EE error is defined as
\begin{equation}
    |\Delta \mathrm{EE}| = |\mathrm{EE_{genie}} - \mathrm{EE}_\mathcal{X}|,
\end{equation}
where $\mathrm{EE_{genie}}$ is the EE obtained under perfect position knowledge and $\mathrm{EE}_\mathcal{X}$ is the EE obtained using the position predicted by model $\mathcal{X}$. The genie in both metrics is a reference, and not an upper bound. As such it does not facilitate as ``optimal beamforming". Measuring both SNR and EE against perfect position knowledge means that everything else (path loss, multipath, noise realization) cancels. The differential is therefore attributable to beam misalignment alone. 

As illustrated in Fig.~\ref{fig:BMP}, both learned models outperform the $k$-NN model, especially near the BS, before the 3 meters. After this region, the gap between the models narrows. At the far-region, all models exhibit negligible difference in quality degradation, with differences in the order of $10^{-1}$~dB. This is expected, as LoS users often experience less degradation due to beam misalignment, even in high-frequency scenarios \cite{Zia2022}.

\begin{table}[h]
\centering
\caption{Link-level energy efficiency at $K=100$ users,
$\Delta t = 10~\mathrm{ms}$.}
\begin{tabular}{lrrrrr}
\toprule
Scheme & SNR [dB] & $P_{\mathrm{inf}}$ [W] & EE w/o $P_{\mathrm{inf}}$
       & EE w/ $P_{\mathrm{inf}}$\\
\midrule
Genie  & $32.57$ & $0.000$ & $3.002\times10^{7}$ & ---\\
CNN    & $32.66$ & $4.392$ & $3.010\times10^{7}$ & $2.738\times10^{7}$\\
CSNN   & $32.63$ & $1.122$ & $3.007\times10^{7}$ & $\mathbf{2.932\times10^{7}}$\\
$9$-NN & $32.51$ & $5.803$ & $2.997\times10^{7}$ & $2.649\times10^{7}$\\
\bottomrule
\label{tab:linkenergy}
\end{tabular}
\end{table}

For the EE we assumed a scenario of $K=100$ users and a beam-update interval of $10$~ms. As calculated from our simulation parameters, the link power budget is divided as $7.5\%$ to the $P_t/\mathrm{PA}_e$ ratio, $22.6\%$ is the $P_\mathrm{co}$, the RF chains amass just a $0.6\%$, while the phase shifters account for the $69.3\%$ of the total consumed energy, excluding the inference consumption. The $1024$ phase shifters consume more than twice the baseline circuit power and nine times the amplified transmit power. This is a direct consequence of the large array that mmWave beamforming requires, and it sets the scale against which $P_\mathrm{inf}$ must be judged.

In Table~\ref{tab:linkenergy} we can observe that the positioning accuracy barely affects the link. The mean SNR spans $0.15$ dB across schemes whose $\mathrm{CE}_{90}$ differs by a factor of $5.4$. Consequently, the EE computed without the inference contribution varies by only $0.4\%$. At the ranges covered by this deployment, the beam footprint is wide relative to the positioning error for the great majority of positions, so even a $0.635$~m error costs almost nothing in received power. The exception is the near-BS region, which is the subject of Fig.~\ref{fig:BMP}.

\begin{figure}
    \centering
    \includegraphics[width=\columnwidth]{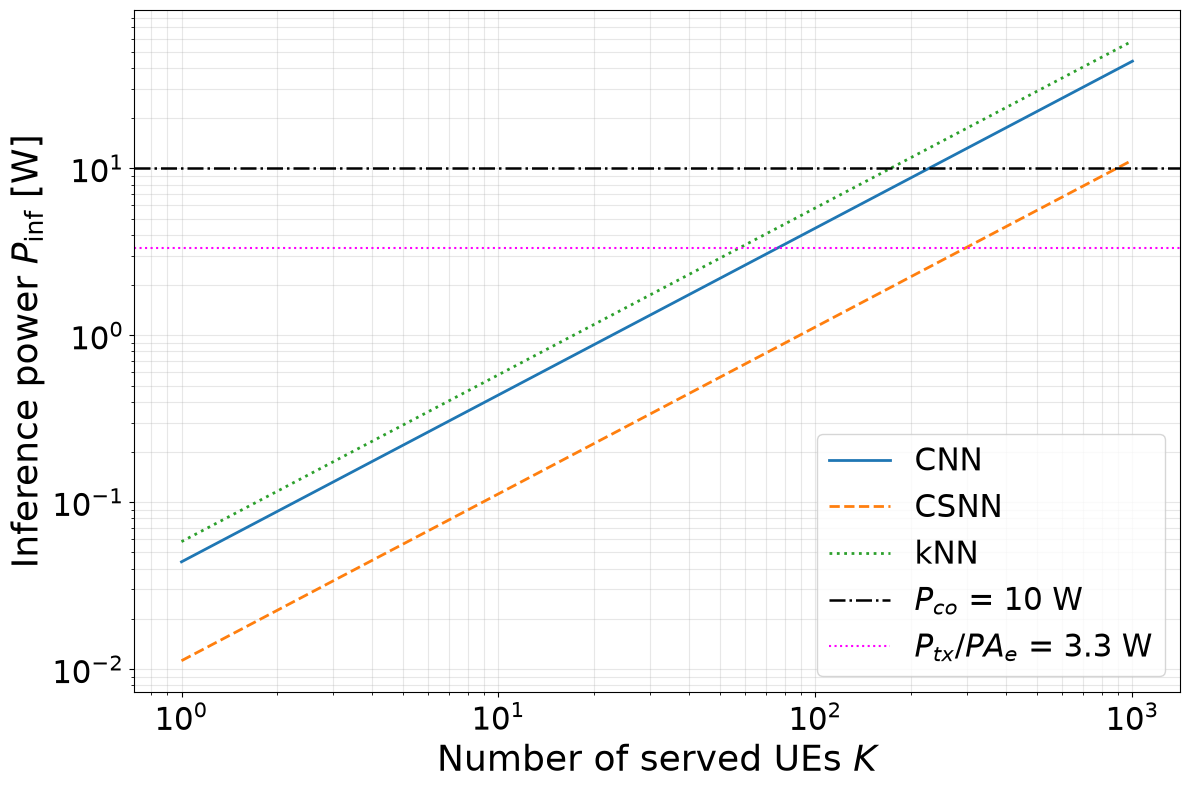}
    \caption{Isolated scaling behavior of inference power for constant update interval $\Delta t$.}
    \label{fig:modelpowerscale}
\end{figure}

Including the inference power consumption, the spread widens to $10\%$ and the ordering inverts; the CSNN attains the highest EE despite having the second-worst positioning accuracy, because its inference costs $3.9$ times less than the CNN's and its SNR penality is $0.03$~dB. The CSNN loses only $2.5\%$ of the genie-aided EE, against $9.0\%$ for the CNN and $11.6\%$ for the $k$-NN. As such, the accuracy that the CNN ``buys" is not worth its energy in our deployment, as the number of users scales. The power scaling of each model is illustrated in Fig.~\ref{fig:modelpowerscale}.

Setting $P_\mathrm{inf} = P_\mathrm{co}$ at $\Delta t=10~\mathrm{ms}$ leads to a number of UEs, $K$, equal to $172$, $228$, and $891$, for the $9$-NN, CNN, and CSNN, respectively. The CSNN sustains roughly four times as many users before its positioning subsystem reaches the baseline circuit power, meaning that it consumes equal energy with the baseline control circuitry. As such, we can conclude that inference is not necessarily an important power component at small $K$, but its importance grows linearly with the number of served users

\begin{figure}
    \centering
    \includegraphics[width=\columnwidth]{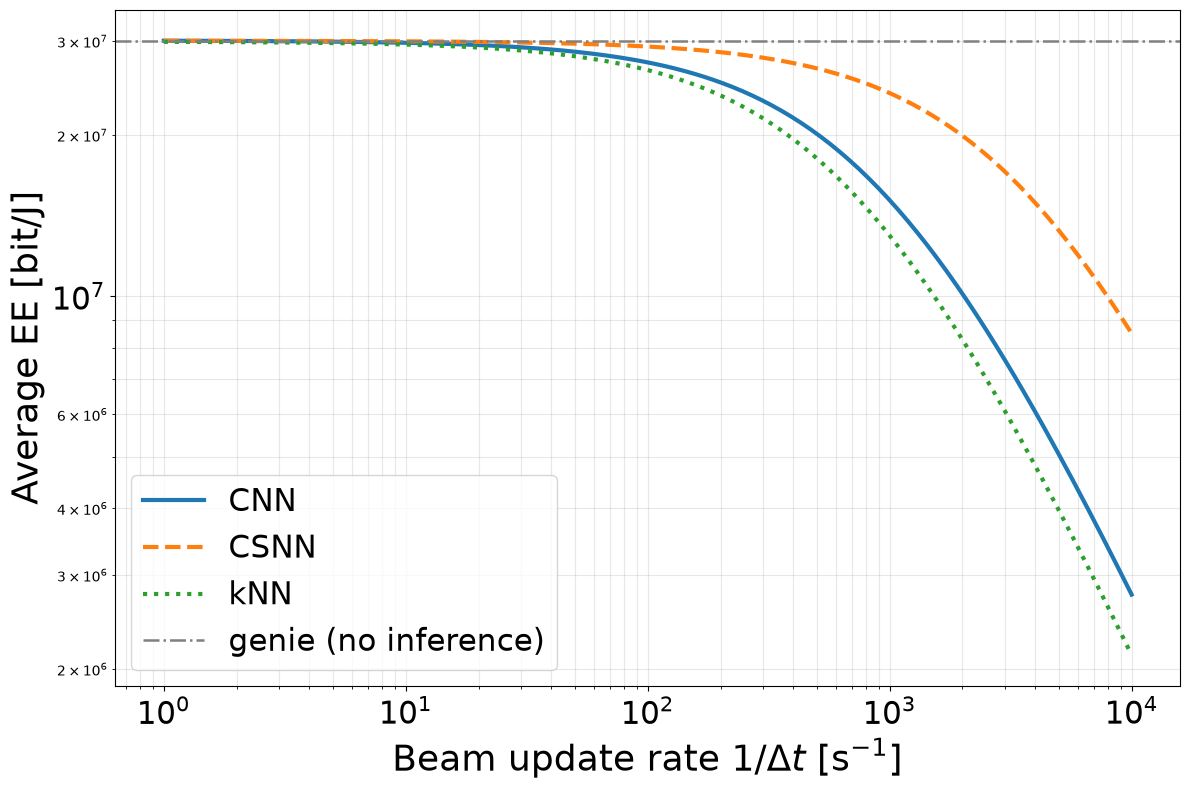}
    \caption{Average link energy efficiency over increasing beam-update rates. The simulation is run for $K=100$ UEs, to produce non-negligible power differentials.}
    \label{fig:linkee}
\end{figure}

Fig.~\ref{fig:linkee} illustrates the EE against the beam-update rate, i.e., the frequency at which a new inference is invoked. The CSNN is above the CNN at every rate, converging only as $1/\Delta t \rightarrow0$ where $P_\mathrm{inf}$ vanishes for all schemes. Increasing the update rate increases the inference power because the positioning model is executed more frequently. Consequently, the energy efficiency decreases with update rate. The degradation is much smaller for CSNN because its inference energy per prediction is substantially lower than that of CNN and $k$-NN. There is no crossing point between the curves. The reason is that the CSNN's SNR deficit is far too small to offset its energy advantage at any positive update rate. This means that the choice of model is not a trade-off in this operating regime, and the cheaper model is preferable throughout.

\section{CONCLUSION}\label{sec:conclusion}

In this paper, we developed an CSNN-based positioning subsystem for a mmWave system. The system leverages CSI obtained from UL transmissions to estimate user positions for downlink beam steering. The proposed CSNN demonstrated positioning performance comparable to that of a conventional CNN while requiring significantly less energy per inference. In particular, the CSNN achieved the lowest energy-per-inference measurement among all models evaluated in this study. From a positioning accuracy perspective, its performance ranked in the middle of the tested models, indicating a trade-off between accuracy and energy efficiency. Moreover, in scatterer poor environment and LoS regions near the BS, small positioning errors produce disproportionately large SNR and EE penalties under narrow-beam DL, whereas rich scattering partially compensates for larger positioning errors. Consequently, the proposed  CSNN is well suited to edge-deployed, low-power inference applications where moderate positioning accuracy is acceptable. 
Future works based on the end-to-end ray-traced  urban dataset and generation pipeline for CSI-based positioning at mmWave band may explore alternative network architectures, layer configurations, and hyperparameter settings, as well as evaluate performance across additional communication scenarios. 

\bibliographystyle{IEEEtran}
\bibliography{IEEEabrv,cleaned}

\end{document}